\documentclass[11pt]{article}
\newcommand{\Comment}[1]{{}}
\usepackage{amsfonts,amsthm,amsmath,amssymb,slashed}
\usepackage[textwidth = 430 pt, textheight = 630 pt]{geometry}
\usepackage{color}

\Comment{\usepackage{color}
\definecolor{MyDarkBlue}{rgb}{0.15,0.15,0.45}
\usepackage[linktocpage=true]{hyperref}
\hypersetup{
colorlinks=true,
citecolor=MyDarkBlue,
linkcolor=MyDarkBlue,
urlcolor=MyDarkBlue,
pdfauthor={Jeff Murugan, Horatiu and Nastase},
pdftitle={A nonabelian particle-vortex duality in gauge theories},
pdfsubject={hep-th}
}

\usepackage[numbers,sort&compress]{natbib}
\usepackage{hypernat}}
\usepackage{graphicx}
\usepackage{cite}

\newcommand\ignore[1]{}
\def\one{{\,\hbox{1\kern-.8mm l}}}

\def\Tr{{\rm Tr\, }}

\def\a{\alpha}

\def\e{\epsilon}

\def\d{\partial}

\def\Tr{\mathop{\rm Tr}\nolimits}

\newcommand{\Cset}{{\,\,{{{^{_{\pmb{\mid}}}}\kern-.45em{\mathrm C}}}}}

\newcommand{\be}{\begin{equation}}
\newcommand{\bea}{\begin{eqnarray}}

\newcommand{\ee}{\end{equation}}
\newcommand{\eea}{\end{eqnarray}}

\begin{document}

\renewcommand{\thefootnote}{\fnsymbol{footnote}}

\makeatletter
\@addtoreset{equation}{section}
\makeatother
\renewcommand{\theequation}{\thesection.\arabic{equation}}

\rightline{}
\rightline{}

\begin{flushright}
QGASLAB-15-09
\end{flushright}

\vspace{10pt}


\begin{center}
{\LARGE \bf{\sc A nonabelian particle-vortex duality in gauge theories}}
\end{center} 
 \vspace{1truecm}
\thispagestyle{empty} \centerline{
{\large \bf {\sc Jeff Murugan${}^{a,}$}}\footnote{E-mail address: \Comment{\href{mailto:jeff@nassp.uct.ac.za}}{\tt jeff@nassp.uct.ac.za}} 
{\bf{\sc and}}
{\large \bf {\sc Horatiu Nastase${}^{b,}$}}\footnote{E-mail address: \Comment{\href{mailto:nastase@ift.unesp.br}}{\tt nastase@ift.unesp.br}} 
                                                           }

\vspace{.5cm}

 
\centerline{{\it ${}^a$
The Laboratory for Quantum Gravity \& Strings, }} \centerline{{\it
Department of Mathematics and Applied Mathematics, }} \centerline{{\it
University of Cape Town, Private Bag, Rondebosch, 7700, South Africa}}

\centerline{{\it ${}^b$ 
Instituto de F\'{i}sica Te\'{o}rica, UNESP-Universidade Estadual Paulista}} \centerline{{\it 
R. Dr. Bento T. Ferraz 271, Bl. II, Sao Paulo 01140-070, SP, Brazil}}

\vspace{1truecm}

\thispagestyle{empty}

\centerline{\sc Abstract}

\vspace{.4truecm}

\begin{center}
\begin{minipage}[c]{380pt}{\noindent We define a nonabelian version of particle-vortex duality, by dimensionally extending usual (1+1)-dimensional 
nonabelian T-duality to (2+1) dimensions. While we will explicitly describe a global $SU(2)$ symmetry, our methods can also be applied to a larger 
group $G$, by gauging an appropriate subgroup. We will exemplify our duality with matter in both adjoint and fundamental representations by 
considering a modification of ${\cal N}=2$ supersymmetric Yang-Mills theory (Seiberg-Witten theory reduced to (2+1) dimensions), and an 
$SU(2)\times U(1)$ color-flavor locked theory that exhibits nonabelian 
vortex solutions. 
}
\end{minipage}
\end{center}

\vspace{.5cm}

\setcounter{page}{0}
\setcounter{tocdepth}{2}

\newpage

\renewcommand{\thefootnote}{\arabic{footnote}}
\setcounter{footnote}{0}

\linespread{1.1}
\parskip 4pt



\section{Introduction}

\noindent
The observation that the vacuum Maxwell equations are invariant under an exchange of electric and magnetic fields is sufficiently 
innocuous that we often teach it to freshman students. At the same time, it is deep enough that it is the direct ancestor of the web 
of low-dimensional dualities shaping our contemporary understanding of quantum field theory. Arguably, apart from this electric-magnetic 
case, the best known, and first formal realization of a duality between two field theories, was Coleman's proof of the (quantum) equivalence 
of the massive Thirring and Sine-Gordon models in (1+1) dimensions \cite{Coleman:1974bu}. Here, in mapping between weak and strong 
coupling physics, the duality exchanges {\it particles} of the one theory with topological {\it solitons} of the other. This precipitated an 
enormous amount of work in the 80's and 90's that culminated in Seiberg and Witten's landmark analysis of ${\cal N}=2$ supersymmetric 
gauge theories in (3+1) dimensions. In this case, the electric-magnetic duality that exchanges particles with monopoles was a crucial 
ingredient in solving for the full low energy effective action \cite{Seiberg:1994rs,Seiberg:1994aj}.\\

\noindent
That said, the bulk of the literature on the subject remains devoted to 1+1 dimensions, with application to integrability and spin chains, 
and 3+1 dimensions relevant to particle physics. The intermediate case of two spatial dimensions has remained largely unexplored. 
In this planar case, the duality acts between particles and the corresponding co-dimension 2 topological defects, {\it vortices}. As such, it
finds application primarily in condensed matter physics, particularly in anyonic superconductivity \cite{Lee:1991jt} and the quantum 
Hall effect \cite{Burgess:2000kj}. However, this particle-vortex duality (or 3D mirror symmetry) is not without its subtleties. In particular, 
while the electric-magnetic duality used to solve for the low energy action of ${\cal N}=2$ supersymmetric gauge theories was perfectly 
well defined for nonabelian theories, a corresponding non-abelian realisation of particle-vortex duality in (2+1) dimensions was not. 
Indeed, even the abelian duality itself was not very well understood \cite{Zee}. 
In \cite{Murugan:2014sfa} a path integral definition was given and embedded into the ${\cal N}=6$ supersymmetric ABJM model
\cite{Aharony:2008ug}, an $SU(N)\times SU(N)$ Chern-Simons matter gauge theory dual to the type IIA superstring in 
$AdS_4\times \mathbb{CP}^3$. In this context, it was shown that abelian particle-vortex duality is holographically dual to 
Maxwell duality in the $AdS_4$ bulk.\\

\noindent
In the prequel to this work \cite{Murugan:2015boa}, we defined a non-abelian version of particle-vortex duality. Our construction was based 
on a generalization of nonabelian T-duality on the (1+1)-dimensional worldsheet of a string, to (2+1) dimensions. This holds for theories 
with a global nonabelian $SU(2)$ invariance, as well as a local (gauge) symmetry. This was possible since, from the path integral point 
of view, abelian T-duality is nothing but a (1+1)-dimensional version of the (2+1)-dimensional particle-vortex duality. In this paper we 
expand on this non-abelian particle-vortex duality. In particular, we generalize it to a theory invariant under an arbitrary group $G$, 
as well as considering matter in a general representation. We illustrate our construction with several concrete examples. In addition 
to elaborating on the semilocal cosmic strings reported in \cite{Murugan:2015boa}, we also apply the duality to a modification of 
the ${\cal N}=2$ gauge theory of Seiberg and Witten, reduced to 2+1 dimensions. Our final example is that of an 
$SU(2)\times U(1)$ gauge theory that exhibits non-abelian color-flavor locked vortices \cite{Auzzi:2003fs}. \\

\noindent
The paper is organized as follows. We set out our notation and formalism in section 2 by reviewing Abelian and non-Abelian T-dualities. 
Section 3 contains the crux of our argument. Here, after a review of the abelian version, we define the non-Abelian duality, as well as several 
generalizations. Section 4 is devoted to several explicit examples that serve to illustrate the workings of the duality, and we conclude 
with a discussion and some future directions in section 5. We include also an appendix in which we make some comments on a generalization to a general matter representation.

\section{T-duality}

\subsection{Abelian T-duality}
In this subsection we give a streamlined review Buscher's path integral derivation of (abelian) T-duality \cite{Buscher:1987qj} in the context of the $(1+1)$-dimensional string worldsheet\footnote{We will, without loss of generality, restrict ourselves to the case of $B_{\mu\nu}=0$ and $\phi = $constant. It won't hurt to set $2\pi\alpha' = 1$ either.}. This is, of course, the quintessential example of duality in the path integral, and the other cases are simply generalizations thereof. \\

\noindent
One begins with a ``master" first order action for the coordinates, $X^i$, not modified by T-duality; a first order set of variables
$V_\mu$; and the Lagrange multipliers $\hat x^0$ that enforce flatness of the connection $V_{\mu}$,
\be
  S_{\rm master}=\int d^2\sigma \left\{\sqrt{\gamma}\gamma^{\mu\nu}[g_{00}V_\mu V_\nu+2g_{0i} V_\mu 
  \d_\nu X^i+g_{ij}\d_\mu X^i \d_\nu X^j] +2\epsilon^{\mu\nu}\hat X^0 \d_\mu V_\nu\right\}.
\ee
Variation of this action with respect to the Lagrange multiplier $\hat X^0$ gives $\epsilon^{\mu\nu}\d_\mu V_\nu=0$. This equation, in turn, is solved by $V_\mu=\d_\mu X^0$. Replacing this $V_\mu$ in the master action takes us back to the usual string action, 
\be
  S_{\rm P}=\int d^2\sigma \left\{\sqrt{\gamma}\gamma^{\mu\nu}[g_{00}\d_\mu X^0 \d_\nu X^0+2g_{0i} \d_\mu X^0\d_\nu   
  X^i+g_{ij}\d_\mu X^i \d_\nu X^j]\right\}. 
\ee
On the other hand, varying instead with respect to $V_\mu$ gives 
\be
  V_\mu=\frac{\hat g_{00}}{\sqrt{\gamma}}\epsilon^{\nu\rho}\gamma_{\rho\mu} \d_\nu \hat X^0-g_{0i}\d_\mu \hat X^i\;,
\ee
where $\hat X^i=X^i$, $\hat g_{00}=1/g_{00}$. Replacing this $V_\mu$ in the master action gives the 
string action in the T-dual coordinates, 
\be
  \widetilde{S}=\int d^2\sigma\left[\sqrt{\gamma}\gamma^{\mu\nu}\hat g_{ab}\d^\mu \hat X^a\d_\nu \hat X^b+\epsilon^{\mu\nu}\hat B_{ab}\d_\mu \hat X^a \d_\nu
\hat X^b\right]\;,
\ee
where $a=(0i)$, and $\hat g_{ab}$ and $\hat B_{ab}$ are the T-dual metric and B-field,
\be
\hat g_{ij}=g_{ij}-\frac{g_{0i}g_{0j}}{g_{00}};\;\;\;
\hat B_{0i}=\frac{g_{0i}}{g_{00}}.
\ee 
The relation between $X^0$ and $\hat X^0$ is found by equating the two formulas for $V_\mu$, 
\be
\d_\mu X^0=\frac{\hat g_{00}}{\sqrt{\gamma}}\epsilon^{\nu\rho}\gamma_{\rho\mu} \d_\nu \hat X^0-g_{0i}\d_\mu \hat X^i=
\frac{\hat g_{00}}{\sqrt{\gamma}}\epsilon_{\rho\mu}\d^\rho \hat X^0-g_{0i}\hat X^i.\label{tdualmap}
\ee
This makes clear the point that, in the absence of $g_{0i}$, T-duality looks like Hodge duality ($*\widetilde{F}=F$) in 2 dimensions.

\subsection{Nonabelian T-duality}

The nonabelian generalization of the above T-duality transformation was first defined in \cite{delaOssa:1992vc}, and only recently modified to include RR-fields in \cite{Sfetsos:2010uq,Itsios:2013wd}. It is a transformation on a compact space, invariant under the action of a general nonabelian group, $G$. For concreteness (and simplicity), we will choose $G=SU(2)$, although our discussion below extends, in principle, to other group manifolds. The metric and B-field on the space are written in terms of the metric on the group manifold. Specifically, in terms of left-invariant one-forms, $L^i$, on the group manifold that satisfy
\be
dL^i=\frac{1}{2}{f^i}_{jk}L^j\wedge L^k\;,
\ee
the metric and $B$-field, respectively, are expressed as
\bea
  ds^2&=&G_{\mu\nu}dx^\mu dx^\nu +2G_{\mu i}dx^\mu L^i+g_{ij}L^iL^j,\cr
  B&=& B_{\mu\nu}dx^\mu \wedge dx^\nu +B_{\mu i}dx^\mu \wedge L^i+\frac{1}{2}b_{ij}L^{i}\wedge L^j.
\eea
Together with the constant dilaton $\phi=\phi_0$ they form an NS-NS string background. For the specific case of $G=SU(2)$, the $L^i$ are the usual $SU(2)$ left-invariant one-forms, 
\bea
  L_1&=& \frac{1}{\sqrt{2}}(-\sin\psi d\theta +\cos\psi \sin \theta  d\phi),\cr
  L_2&=&\frac{1}{\sqrt{2}}(\cos\psi d\theta +\sin\psi\sin\theta d\phi),\\
  L_3&=&\frac{1}{\sqrt{2}}(d\psi+\cos\theta d\phi)\;,\nonumber
\eea
written in terms of the Euler angles $(\theta,\phi,\psi)$, which range over $0\leq \theta\leq \pi$, $0\leq \phi\leq 2\pi$, $0\leq \psi\leq 4\pi$. Using the normalized Pauli matrices, $t^i=\tau^i/\sqrt{2}$, as generators, with $\Tr(t^it^j)=\delta^{ij}$, and the group element
\be
  g=e^{\frac{i\phi \tau_3}{2}}e^{\frac{i\theta\tau_2}{2}}e^{\frac{i\psi\tau_3}{2}},
\ee
understood as a field on the string worldsheet, the one-forms can be rewritten in world-sheet light-cone coordinates as
\be
L^i_\pm=-i\Tr(t^ig^{-1}\d_\pm g).
\ee
Of course, the above formula for the $L^{i}$ holds more generally for $t^i$ taken to be the normalized generators of some other group $G$ and $g$ an element of the group. Note also that, in this construction, $g$ is complex while the $L^{i}$ are all real.
Defining
\bea
  Q_{\mu\nu}&=&G_{\mu\nu}+B_{\mu\nu},\;\;\;
  Q_{\mu i}=G_{\mu i}+B_{\mu i},\cr
  Q_{i\mu}&=& G_{i\mu}+B_{i\mu},\;\;\;
  E_{ij}=g_{ij}+b_{ij}\;,
\eea
the Polyakov string worldsheet action is written in a global $SU(2)$-invariant form as
\be
  S=\int d^2\sigma \left[Q_{\mu\nu}\d_+X^\mu \d_-X^\nu +Q_{\mu i}\d_+X^\mu L_-^i
  +Q_{i\mu} L_+^i \d_-X^\nu +E_{ij}L_+^i L_-^j\right].\label{origstring}
\ee
Generalizing (\ref{tdualmap}), we want to perform a nonabelian T-duality transformation 
on the coordinate one-forms $L^{i}$ of the $G$-invariant manifold. Toward that end, we gauge the global $G$-invariance 
($SU(2)$ in this particular case) as usual by introducing a gauge field $A$, and making derivatives covariant so that, 
\be
  \d_\pm g\rightarrow D_\pm g=\d_\pm g-A_\pm g\;.
\ee
This leads to the replacement 
\be
  \tilde L_\pm ^i=-i\Tr[t^i g^{-1}D_\pm g],
\ee
in the string action. As in the abelian case, the gauge field $A$ that we introduce, the analog of $V_\mu$ above, must be auxiliary in order not to introduce any extra degrees of freedom. This is accomplished by imposing its triviality, $F_{\mu\nu}=0$, as a constraint, through a Lagrange multiplier term, 
\be
  -i\Tr[vF_{+-}]=-i\epsilon^{\mu\nu}\Tr[vF_{\mu\nu}],
\ee
 in the action. Here $v$ is an adjoint vector of $G$, and the nontrivial field strength components are
\be
  F_{+-}=\d_+A_--\d_-A_+-[A_+,A_-].
\ee
In this way, we obtain a master action for the nonabelian T-duality, 
\be
S=\int d^2\sigma \left[Q_{\mu\nu}\d_+X^\mu \d_-X^\nu +Q_{\mu i}\d_+X^\mu \tilde L_-^i
+Q_{i\mu} \tilde L_+^i \d_-X^\nu +E_{ij}\tilde L_+^i \tilde L_-^j-i\Tr[v F_{+-}]\right].
\ee
Integrating out the Lagrange multiplier $v$ leads to $F_{+-}=0$, which means that the gauge field is trivial, modulo any topological issues, and substituting $A=0$ we go back to the original action (\ref{origstring}). \\

\noindent
Integrating out the gauge field $A_\pm$ instead, obtains the string action in the nonabelian T-dual background. There is however one important caveat; now we have too many degrees of freedom. Essentially, this is because we have retained the original coordinates $L_i$ or, more generally, the coordinates $g$ on the group manifold. Before solving for $A_\pm$ in terms of the 
$v^i$ that will play the role of the dual coordinates, we need to first gauge fix the local $G$ symmetry. One simple gauge fixing 
choice is to set $g=1$, leading to the replacement
\be
  L^i_\pm \rightarrow i\Tr[t^iA_\pm]=iA^i_\pm.
\ee
Then, partially integrating the Lagrange multiplier term sets
\be
  -i\int d^2\sigma \Tr[vF_{+-}]=\int d^2\sigma \left\{ \Tr[+i(\d_+v)A_--i(\d_-v)A_+]-A_+fA_-\right\}\;,
\ee
where 
\bea
  &&A_+\,f\,A_-\equiv A_+^i\,f_{ij}\,A_-^j\cr
  &&f_{ij}\equiv {f_{ij}}^{k}\,v_{k}\;.
\eea
Then, replacing the above and setting $L^{i}=iA^i$ in the master action, we obtain
\bea
  S&=&\int d^2\sigma\left[Q_{\mu\nu}\,\d_+ X^\mu \d_- X^\nu+Q_{\mu i}\,\d_+ X^\mu(+iA_-^i)
  +Q_{i\mu}\,\d_-X^\mu (+iA_+^i)\right.\cr
  &&\left.+E_{ij}(iA_+^i)(iA_-^j)+i\d_+v_i\, A_-^i-i\d_-v_i\, A_+^i-A_+^i\,f_{ij}\,A_-^j\right].
\eea
To arrive at the dual action, we need to eliminate the auxiliary gauge fields. To this end, varying the above action with respect to 
$A_+^i$ and, respectively, $A_-^j$ gives
\begin{eqnarray}
  f_{ij}A_-^j&=&-i\d_-v_i-E_{ij}A_-^j+iQ_{\mu i} \d_- X^\mu\;,\nonumber\\
  A_+^i f_{ij}&=&+i\d_+v_j-E_{ij}A_+^i+iQ_{\mu j}\d_+X^\mu,
\end{eqnarray}
which can be solved to give
\bea
A_-^i&=&-iM_{ij}^{-1}(\d_- v_j-Q_{j\mu}\d_- X^\mu)\cr
A_+^i&=&+iM_{ij}^{-1}(\d_+ v_j+Q_{\mu j}\d_+ X^\mu)\;,
\eea
where we have defined $M_{ij}\equiv E_{ij}+f_{ij}$. Finally, substituting $A^i_\pm$ back into the master action gives the dual action
\be
S_{\rm dual}=\int d^2\sigma[Q_{\mu\nu}\d_+X^\mu \d_- X^\nu
+(\d_+v_i+Q_{\mu i}\d_+X^\mu)M^{-1}_{ij}(\d_-v_j-Q_{j\mu}\d_- X^\mu)].
\ee
At the quantum, one-loop determinant, level, it should be noted that the dilaton is modified in the usual way, as 
$\Phi(x,v)=\Phi(x)-\frac{1}{2}\ln(\det M)$.

\section{Particle-vortex duality in 3 dimensions}

Having set up the nonabelian T-duality procedure, we are now in a position to lift it to three dimensions where it will become particle-vortex duality. Before doing so, however, it will be useful to remind the reader of the more familiar abelian particle-vortex duality\footnote{For a more `textbook' introduction, the reader is referred to \cite{Zee}} as formulated in \cite{Murugan:2014sfa}.

\subsection{Abelian particle-vortex duality}

The story here starts with a general action for a complex scalar coupled to an abelian gauge field,
\be
  S=\int d^3x \left[-\frac{1}{2}|D_\mu\Phi|^2-V(|\Phi|)-\frac{1}{4}F_{\mu\nu}^2\right]\;,
\ee
where, as usual, $F_{\mu\nu}=\d_\mu a_\nu-\d_\nu a_\mu$ and $D_\mu\Phi=\d_\mu\Phi -iea_\mu\Phi$. With an appropriate choice of symmetry-breaking potential, the theory admits vortex solutions, $(\Phi,a_\mu)$, {\it i.e.} isolated zeros of $\Phi$ with nontrivial monodromy. To rewrite the on-shell action in terms of the vortex variables, we first express the scalar as $\Phi=\Phi_0 e^{i\a}$ where, in the presence of vortices the argument splits into $\a=\a_{\rm smooth}+\a_{\rm vortex}$ with $\alpha_{\rm vortex}$ encoding the $2\pi n$ phase change that comes with a counter-clockwise circling of the vortex. Consequently 
\bea
  S&=&-\frac{1}{2}\int d^3x\left[(\d_\mu\Phi_0)^2+(\d_\mu \a_{\rm smooth}+\d_\mu \a_{\rm vortex}+ea_\mu)^2\Phi_0^2+  
  \right]\cr
  &&-\int d^3x \left[V(\Phi_0)+\frac{1}{4}F_{\mu\nu}^2\right].
\eea
It is not difficult to see that this can be obtained from the `master' action 
\bea
  S_{\rm master}&=&\int d^3x\left[-\frac{1}{2}(\d_\mu\Phi_0)^2-\frac{1}{2}
  (\lambda_{\mu,{\rm smooth}}+\lambda_{\mu,{\rm vortex}}+ea_\mu)^2\Phi_0^2
  +\epsilon^{\mu\nu\rho} b_\mu \d_\nu \lambda_\rho\right.\cr
  &&\left.-V(\Phi_0)-\frac{1}{4}F_{\mu\nu}^2\right],
\eea
by varying with respect to the Lagrange multiplier $b_\mu$, solving the resulting equation of motion for 
$\lambda_\mu=\d_\mu\a$ and subtituting into the master action. Varying instead with respect to 
$\lambda_{\mu,{\rm smooth}}$, gives 
\be
  (\lambda_\mu+ea_\mu)\Phi_0^2=-\epsilon^{\mu\nu\rho}\d_\nu b_\rho.
  \label{lambdamu}
\ee
Substituting this $\lambda_\mu$ back into the master action, gives the dual action
\be
  S_{\rm dual}=\int d^3x \left[-\frac{(f^b_{\mu\nu})^2}{4\Phi_0^2}-\frac{1}{2}(\d_\mu \Phi_0)^2-\epsilon^{\mu\nu\rho}b_  
  \mu \d_\nu a_\rho
  -\frac{2\pi}{e}b_\mu j^\mu_{\rm vortex}-V(\Phi_0)-\frac{1}{4}F_{\mu\nu}^2\right].
\ee
Two points worth noting, are that under the particle-vortex duality, the scalar current 
\be
  j_\mu=-ie(\Phi^\dagger \d_\mu \Phi-\Phi \d_\mu \Phi^\dagger)=e\Phi_0^2\d_\mu\a\;,
\ee
associated to the the complex field $\Phi$ is exchanged with the vortex current
\be
  j^\mu_{\rm vortex}=\frac{1}{2\pi }\epsilon^{\mu\nu\rho}\d_\nu\d_\rho\a=\frac{1}{2\pi e\Phi_0^2}
  \epsilon^{\mu\nu\rho}\d_\nu j_\rho\;,
\ee
and secondly, the real phase, $\a$, associated to the vortex field is interchanged with the gauge field $b_\mu$, while the original 
gauge field $a_\mu$ remains unaffected. 

\subsection{Nonabelian particle-vortex duality}
We now have all the ingredients to formulate our nonabelian particle-vortex duality by lifting the non-abelian T-duality prescription to  
three dimensions, but writing it in the familiar language of abelian particle-vortex duality. As this is really the crux of our argument, we will make every effort to be pedagogical in our treatment.

\noindent
To begin, we consider the real variables $\Phi_0^k$ and $L_\mu^i$ that are the nonabelian generalizations of 
$\Phi_0$ and $\lambda_\mu$ respectively. At this point, the indices $i$ and $k$ are not restricted in how they transform. Taking the matrices $t^i$ in the Lie algebra of $G$, we construct the {\em real} variables
\be
  L_\mu^i=-i\Tr[t^i g^{-1}\d_\mu g]\;,
\ee
where, again, $g\in G$ is a group element and $\mu=1,2,3$ are spacetime indices in 2+1 dimensions. 
Next, we want to define an action for $L_\mu^i$ and $\Phi_0^k$. In close analogy with the two-dimensional case, we need to 
introduce analogs of the $Q$ and $E$ matrices which, in the present case, we will call $Q_{kl}$, $Q_{ki}$ and $E_{ij}$. 
However, we find that the introduction of general $Q_{kl}$ and $Q_{ki}$ comes at the cost of a loss of a simple interpretation, and consequently we first restrict our attention to the case of $Q_{ki}=0$ and $Q_{kl}=\delta_{kl}$. In this language then,
\be
  S_{\rm original}=\int d^3x \left[-\frac{1}{2}(\d_\mu\Phi_0^k)^2-\frac{1}{2}(\Phi_0^k)^2g^{\mu\nu}L_\mu^i L_\nu^j
  E_{ij}\right].\label{original3d}
\ee
To write a master action for the duality, we gauge a subgroup $G$, by introducing gauge fields $A_\mu^i$
and replacing normal derivatives with covariant derivatives $\d_\mu g\rightarrow D_\mu g=\d_\mu g-A_\mu g$ and 
$L_\mu^i$ with $\tilde L_\mu^i=-i\Tr[t^i g^{-1}D_\mu g]$. The ungauged theory is recovered by imposing a vanishing of the field strength, $F_{\mu\nu}=\d_\mu A_\nu -\d_\nu A_\mu -[A_\mu,A_\nu]$, with Lagrange multipliers $v^i$. Note that, since we gauge only a subgroup of $G$, some of the $A^i$, with their associated $v^i$, are taken to be zero. Putting this all together gives the master action 
\be
  S_{\rm master}=\int d^3x\, \left[-\frac{1}{2}(\d_\mu \Phi_0^k)^2-\frac{1}{2}(\Phi_0^k)^2g^{\mu\nu}\tilde L_\mu^i \tilde L_  
  \nu^j E_{ij}+\epsilon^{\mu\nu\rho}v_\mu^i F_{\nu\rho}^i\right]\;,\label{master3d}
\ee
for the duality. Indeed, varying with respect to $v^i$ leads to $F_{\mu\nu}^i=0$, and the gauge field can be taken to be trivial. $\tilde L_\mu^i$ is then replaced with $L_\mu^i$ leading to the original action (\ref{original3d}). If, instead, we gauge fix $g=1$ and eliminate the corresponding gauge-fixed $A_\mu^i$ from its equation of motion
\be
  \tilde L_\mu^i\rightarrow i\Tr[t^i A_\mu]=iA_\mu^i\;.
\ee
After a partial integration of the Lagrange multiplier term
\be
  \int d^3x\,  \epsilon^{\mu\nu\rho}\, v_\mu^i F_{\nu\rho}^i=
  \int d^3x\, \epsilon^{\mu\nu\rho}\left[(\d_\mu v_\nu^i)A_\rho^i-(\d_\nu v_\mu^i)A_\rho ^i
  +A_\mu^i\, f_{\nu ij}\,A_\rho^j\right]\;,\nonumber
\ee
where $ f_{\nu ij}\equiv f_{ijk}\,v_\nu^k$. This leads to a purely algebraic action for $A_\mu^i$, which can be explicitly eliminated through its equation of motion,
\be
  [(\Phi_0^k)^2g^{\mu\rho}E_{ij}+2\epsilon^{\mu\nu\rho}f_{\nu ij}]A_\rho^j=-\epsilon^{\mu\nu\rho}
  (\d_\nu v_{\rho i}-\d_\rho v_{\nu i}).
\ee
This is solved by 
\be
A_\mu^{i}= -{M^{ij}}_{\mu\rho}\,V_j^\rho\;,
\ee
with $M^{ij}_{\mu\nu}\equiv (M^{-1})^{\mu\nu}_{ij}$ and $V^{\mu}_{i}$ defined through
\bea
M_{ij}^{\mu\rho}&\equiv& [(\Phi_0^k)^2g^{\mu\rho}E_{ij}+2\epsilon^{\mu\nu\rho}f_{\nu ij}]\cr
V_i^\mu&\equiv & \epsilon^{\mu\nu\rho}(\d_\nu v_{\rho i}-\d_\rho v_{\nu i}).\label{mandv}
\eea
Substituting the $A^{i}_\mu$ back into (\ref{master3d}) yields the dual
\bea
S_{\rm dual}&=& \int d^3x \left[-\frac{1}{2}(\d_\mu\Phi_0^k)^2+\frac{1}{2}A_\mu^i M_{ij}^{\mu\rho}A_\rho^j+A_\mu^i V_i^\mu\right]\cr
&=&-\frac{1}{2}\int d^3x\, \left[V_i^\mu M^{-1}_{ij}V_j^\rho +(\d_\mu \Phi_0^k)^2\right].\label{dual3d}
\eea
Notice that the duality amounts to the exchange of $L_\mu^i$ with $v_{i\mu}$, related by
\be
  L_\mu^i\leftrightarrow \tilde L_\mu^i=-M_{\mu\rho}^{ij}\,\epsilon^{\rho\nu\sigma}(\d_\rho v_{\sigma j}-
  \d_\sigma v_{\rho j}).\label{nonabduality}
\ee
To interpret this as the action of the particle-vortex dual of the original theory, we still need to couple the theory to a nontrivial external gauge field, find a system that admits vortices, and couple the action to the vortex current. The external gauge field must reside in the complement of the symmetry group $G$ with respect to the directions on which the T-duality was carried out, since we cannot gauge those directions twice. 

\subsubsection{Standard kinetic terms}
In order to develop some intuition for how the nonabelian duality works, we would like to mirror as closely as possible its abelian
precursor. There, we traded in $\Phi_0$ and $\lambda_\mu=\d_\mu \a$ for the single field $\Phi=\Phi_0 e^{i\a}$ and we would like to be able to do the same with $\Phi_0^k$ and $L_\mu^i$. One simple case is to take $\Phi_0^k$ in the adjoint representation, but not necessarily of the same group. In the case of $L_\mu^i\in G=SU(2)$, for example, $\Phi_0^k$ could be a (real) adjoint triplet, though not necessarily of the same $SU(2)$ group.
We can then define the complex field
\be
  \Phi=\Phi_0^a\,T_a\otimes \exp\left(i\int dx^\mu L_\mu^i F_i^A \widetilde{T}_{A}\right)\;,\label{adjointansatz}
\ee
where $T_a$ and $\widetilde{T}_A$, with $A=1,...,N$,
are adjoint matrices transforming under {\it a priori} different groups. Though not necessary, the indices $A$ and $i$ can 
both be in the same group $G$. The $F_i^A$ are fixed coefficients from which we construct $E_{ij}$.\\ 

\noindent
If we normalize the generators as $\displaystyle \Tr[T_a T_b]=\delta_{ab};\;\;\;\Tr[\widetilde{T}_A\widetilde{T}_B]=\delta_{AB}$ then $\displaystyle \Tr[(T_a\otimes \widetilde{T}_A)(T_b\otimes \widetilde{T}_B)]=\delta_{AB}\delta_{ab}$, and the canonical kinetic term for the field $\Phi$ can be written as
\be
\Tr|\d_\mu\Phi|^2=\Tr\left[|T_a\otimes \widetilde{T}_A(\d_\mu \Phi_0^a \one_A+i\Phi_0^a L_\mu ^i F_i^A)|^2\right]
=N[(\d_\mu\Phi_0^a)^2+(\Phi_0^a)^2L_\mu^i L_\mu ^j \, E_{ij}]\;,
\ee
where we have defined $N\, E_{ij}\equiv F_i^A F_j^A$. Apart from the overall factor of $N$ then, this is precisely the  kinetic term considered in (\ref{original3d}). At this point, there are several ways we can implement this construction:
\begin{itemize}
  \item Arguably, the simplest is to take $\widetilde{T}_A$ to be trivial, {\it i.e.} a singlet. Then, the ansatz for $\Phi$ is
  \be
    \Phi=\Phi^a T_a;\;\;\; \Phi^{a}=\Phi_0^a \exp\left(i \int dx^\mu L_\mu^i F_i\right),\label{phiasimple}
  \ee
  with a corresponding scalar kinetic term 
  \be
    \sum_{a}|\d_\mu \Phi^{a}|^2=(\d_\mu\Phi_0^a)^2+(\Phi_0^a)^2 L_\mu^i L_\nu^jg^{\mu\nu}F_iF_j\;,
  \ee
  This is clearly the same as in (\ref{original3d}), except with $E_{ij}=F_i F_j$ now separable. Note that, in this case, according to 
  our ansatz, there is only one phase multiplying all the $\Phi^a$. This phase is written in terms of several 
  $L_\mu^i$, labelled by the index $i$ that ranges over the whole global symmetry group $G$. Since the phase encodes just 
  one real degree of freedom, we could well have written $L_\mu^i F_i\equiv \bar{L}_{\mu}$. However, for our purposes of
  dualizing a subgroup of $G$, it will be more useful to leave the dependence on the full global symmetry manifest. 

  \item Another possibility is to take $a$ to be an index in another (non-adjoint) representation, while $\widetilde{T}_A$ is again    
  trivial. We can still write the ansatz as (\ref{phiasimple}), only {\em without} $\Phi=\Phi^a T_a$. Here again, the kinetic term 
  is of the form (\ref{original3d}). 

  \item Yet another choice is to have $T_a$ trivial and the $\widetilde{T}_A$ matrices diagonal. In this case the ansatz 
  takes the form
  \be
    \Phi=\begin{pmatrix}\Phi_0 e^{i\theta_1}&&&\\&\Phi_0 e^{i\theta_2}&&\\&&...&\\&&&\Phi_0 e^{i\theta_n}\end{pmatrix}.
  \ee
  It is a particular case of the diagonal ansatz
  \be
    \Phi=\begin{pmatrix}\Phi_1 e^{i\theta_1}&&&\\&\Phi_2 e^{i\theta_2}&&\\&&...&\\&&&\Phi_n e^{i\theta_n}\end{pmatrix}\;,
  \ee
  often considered in the nonabelian vortex literature. The case of a general representation will be studied in a later section.
\end{itemize}

\subsubsection{Coupling to an external gauge field and vortices}
We are now free to add a potential depending only on $\Phi_0^a$ to this action without affecting the duality. We are also free to couple the scalar field to an external gauge field, $a_\mu$, with kinetic term, $+\frac{1}{4}\Tr[f_{\mu\nu}^2]$. 
As discussed earlier, such an external gauge field, $a_\mu=a_\mu^m T_m$, must live in an interal direction not gauged during the 
duality procedure, a statement that amounts to $\Tr[A_\mu T_m]=0$. The minimal coupling to the gauge field, in terms of the variables $\Phi_0^a$ and $L_\mu^i$, would simply be to make the derivative in $\tilde L_\mu^i$ covariant also with respect to $a_\mu$. This would necessitate replacing $\tilde L_\mu^i$ by 
\be
  \tilde{\tilde L}_\mu^i=-i\Tr[t^i g^{-1}(\d_\mu-i(A_\mu^i T_i+a_\mu^mT_m)) g]\;,
\ee
and, consequently, making {\em all the directions in the symmetry group manifold gauge covariant}. However, since we are looking for vortex solutions, whose topology is usually determined from the boundary condition that $D_\mu\Phi\rightarrow 0$ at infinity, 
it is more useful to define a covariant derivative involving $a_\mu$ that acts on the whole field $\Phi$ instead. 

\noindent
In the case of $\Phi$ obtained from the tensor product of adjoint representations (\ref{adjointansatz}), the action of the partial  derivative on $\Phi$ is given by
\be
  \d_\mu \Phi=(T_a\d_\mu \Phi_0^a +T_a\otimes \widetilde{T}_A\, i\Phi_0^a L_\mu^i F_i^A)\one\otimes e^{i\int dx^\mu 
  L_\mu^i F_i^A \widetilde{T}_A}\;.
\ee
Consequently, the covariant derivative {\em with respect to $a_\mu$} acts as
\bea
  D_\mu\Phi&=&\left(T_a\d_\mu \Phi_0^a +T_a\otimes \widetilde{T}_A\,i\Phi_0^a L_\mu^i F_i^A +
  T_a\Phi_0^a\otimes \left[a_\mu^m\widetilde{T}_m,e^{i\int dx^\mu L_\mu^i F_i^A \widetilde{T}_A}\right]\right.\cr
  &&\left.\times e^{-i\int dx^\mu L_\mu^k F_k^A \widetilde{T}_A}\right)
  \one\otimes e^{i\int dx^\mu L_\mu^j F_j^A \widetilde{T}_A}.
\eea
Comparing the two expressions reveals that, in effect, the $L_\mu^i$ field is replaced by
\be
  L_\mu ^i F_i^A\rightarrow L_\mu^i F_i^A + ia_{\mu}^{m}\int\! dx^{\nu}\, L_\nu^i F_i^B {f_{Bm}}^A + {\cal O}((L_\nu^j)^2)\;,
\ee
to first order. The ${f_{Bm}}^C$ are nothing but the structure constants in the commutator $[T_B,T_m]={f_{Bm}}^C T_C$. 
The case when $T_m$ is a singlet, {\it i.e.} when we gauge a $U(1)$ subgroup of $G$, must be treated separately since there 
${f_{Bm}}^C=0$. In that case, however, the covariant derivative is the usual $U(1)$ covariant one, $D_\mu\Phi=(\d_\mu -ia_\mu)\Phi$. It is easily verified that, if the identity $\one$ corresponding to the $U(1)$ direction is among the $t^i$, say $t^0=\one$, and $F_0=1$, the two definitions coincide, since
\be
  \Tr|D_\mu\Phi|^2=(\d_\mu\Phi^a)^2+\tilde L_\mu^i \tilde L_\mu^j E_{ij}(\Phi_0^a)^2\;,
\ee
with $E_{ij}=F_iF_j$ and $\tilde L_\mu^i=\Tr[t^i g^{-1}D_\mu g]$. If $F_0\neq 1$, the two definitions have gauge fields, $a_\mu$, that differ by a factor of $F_0$. 

\noindent
Having coupled the scalar to an external gauge field, we can now specify a vortex ansatz that generalizes the usual abelian-Higgs one. We will restrict our attention to axially symmetric solutions and set the field modulus $\Phi_0^a=\Phi_0^a(r)$, with $r$ the radial direction on the 2-plane. For a vortex phase that winds $N_{A}$ times around the polar angle $\theta$, we write 
\be
  e^{i\int dx^\mu L_\mu^i F_i^A} =e^{iN_A\theta}\;.\label{vortex}
\ee
In the $U(1)$ case, it will suffice to erase the index $A$ in the vortex phase above. The condition $D_\mu\Phi\rightarrow 0$ as $r\rightarrow \infty$ implies, as usual, both a finite total kinetic energy, as well as a nontrivial topological charge for the field configuration, {\it i.e.} a quantized $\oint a_{\theta} d\theta$. With this ansatz, in specific cases and with specific choice of the potential $V(\Phi)$, we can check for the existence of actual vortices. In the next section, we will do just that and consider three specific cases where the known topological solutions can be put in the form implied by our vortex ansatz. For the rest of this section though, we will assume such a solution exists. \\

\noindent
The final step in defining the duality is to introduce a vortex current into the action, and to check that it is, in fact, dual 
to the particle current. We will take as a guide the abelian case, where the phase variable $\a$ is split into $\a_{\rm smooth}+\a_{\rm vortex}$, and with $\a_{\rm vortex}$ encoding the topological charge through $\a_{\rm vortex}=N\theta$ for $N$ coincident vortices. The analogous statement in the nonabelian case would be the splitting of $L_\mu^i$ into 
$L_\mu^i +L_{\mu,{\rm vortex}}^i$ or, in the gauge $g=1$, using $L_\mu^i= iA_\mu^i+L_{\mu,{\rm vortex}}^i$. However, it will prove more convenient to instead consider $A_\mu^i\rightarrow A^i_{\mu, {\rm smooth}}+A^i_{\mu,{\rm vortex}}$
in the gauge fixed action as the analog of splitting $\lambda_\mu\rightarrow \lambda_{\mu,{\rm smooth}}+\lambda_{\mu,{\rm vortex}}$ in the abelian case. Then, substituting into the gauge-fixed form of (\ref{master3d}), and varying with respect to 
$A^i_{\mu, {\rm smooth}}$, we obtain
\be
  A^i_{\mu, {\rm smooth}}+A^i_{\mu,{\rm vortex}}=-{M^{-1}_{ij}}^{\mu\rho}V_j^\rho\;,
\ee
which is the analog of (\ref{lambdamu}). Substituting back into the gauge-fixed form of the master action (\ref{master3d}), 
gives again the dual action (\ref{dual3d}), except with an extra vortex current term coming from the part of the Lagrange multiplier term, $\epsilon^{\mu\nu\rho}v_\mu^iF_{\nu\rho}^i$, linear in $A_\mu^i$, 
\be
  \epsilon^{\mu\nu\rho}v_\mu^i (\d_\nu A^i_{\rho, {\rm vortex}}-\d_\rho A^i_{\nu,{\rm vortex}})\equiv v_\mu^i\, 
  j^{\mu\, i}_{\rm vortex}\;.\label{duality}
\ee
Our ansatz, (\ref{vortex}), implies that $A^i_{\rho,{\rm vortex}}\propto \d_\rho \theta$, so that this vortex current vanishes everywhere except at the location of the vortex. This is consistent with the corresponding abelian current. More precisely, 
\be
  A_{\mu,{\rm vortex}}^i F_i^A= L_{\mu,{\rm vortex}}^iF_i^A=N^A\d_\mu\theta=\d_\mu \a^A\;,
\ee
where $\a^A$ is a phase for the field with index $A$. We can then define
\be
  j_\mu^A=-i(\Phi^{\dagger}_A\d_\mu\Phi^A-\Phi^A\d_\mu \Phi_A^\dagger)=2(\Phi_0)^2\d_\mu\a^A;,
\ee
with no sum over $A$ and constant field vev, $\Phi_{0}$.  Consequently, $\displaystyle A_{\mu,{\rm vortex}}^i =\frac{j_\mu^A}{2\Phi_0^2}F^A_jE^{-1}_{ij}$ which, in turn, implies that
\begin{eqnarray}
  j^{\mu\,i}_{\rm vortex}= \frac{1}{2\Phi_0^2}\epsilon^{\mu\nu\rho}\d_\nu(j_\rho^A F^A_j E^{-1}_{ij})\;.
\end{eqnarray}
This defines the duality relation between the nonabelian vortex and particle currents, in an analogous way to the abelian case. 

\subsection{Adding $Q$'s}
Up until now, in the interests of building on the intuition afforded us by the abelian particle-vortex duality, we have 
set $Q_{ab}=0=Q_{ai}$, essentially so that we could implement the duality on a theory with a standard kinetic term.
Abandoning this constraint, allows us to generalize the two-dimensional nonabelian T-duality in a minimal but interesting way. 
To this end, consider the action
\be
  S_{\rm original}=\int d^3x\left[-\frac{1}{2}\d_\mu\Phi_0^a g^{\mu\nu} \d_\nu \Phi_0^b Q_{ab}-\frac{1}{2}
  (\Phi_0^a)^2 g^{\mu\nu}\tilde L_\mu^i \tilde L_\nu^j E_{ij} - 
  g^{\mu\nu}(\d_\mu \Phi_0^a)\tilde L_\nu^i Q_{ai}\right],
\ee
where, as before, we define $E_{ij}=F_i^A F_j^A$. Now, however, we want also to write $Q_{ai}\equiv Q_{abi}\Phi_0^b$.
Then, the master action for the 3-dimensional nonabelian duality is 
\bea
  S_{\rm master}&=&\int d^3x\; \left[-\frac{1}{2}(\d_\mu\Phi_o^a)\d_\nu\Phi_0^b\, g^{\mu\nu}\, Q_{ab}
  -\frac{1}{2}(\Phi_0^a)^2 g^{\mu\nu}\tilde L_\mu^i \tilde L_\nu^j E_{ij}\right.\cr
  &&\left.-g^{\mu\nu}(\d_\mu\Phi_0^a)\tilde L_\nu^i\,\Phi_0^v\,Q_{abi}
  +\epsilon^{\mu\nu\rho}\, v_\mu^i  F_{\nu\rho}^j\right].
\eea
As before, solving for the Lagrange multipliers $v_\mu^i$, allows us to set $A^i_{\mu}=0$, reducing $\tilde L_\mu^i$ to $L_\mu^i$, and taking us back to the original action. If, instead, we gauge fix $g=1$, effectively replacing $\tilde L_\mu^i$ by $iA_\mu^i$, the variation with respect to $A_\mu^i$ gives 
\bea
  \left[(\Phi_0^a)^2g^{\mu\rho}E_{ij}+2\epsilon^{\mu\nu\rho}f_{\nu ij}\right]A_\rho^j &=&
  -\epsilon^{\mu\nu\rho}(\d_\nu v_{\rho i}-\d_\rho v_{\nu i})+ig^{\mu\rho}(\d_\rho \Phi_0^a)\Phi_0^b Q_{abi}
  \nonumber\\
  &\equiv& -V^\mu_i + iW_\mu^i\;,
\eea
where $V_i^\mu$ is the same as in the previous subsection and $W_\mu^i$ comes from the last term. This equation is solved by
$A_\mu=-{M^{-1}}_{ij}^{\mu\rho}(V_j^\rho -iW_j^\rho)$. Finally, substituting back into the master action, gives the 
dual action,
\bea
  S_{\rm dual}&=&-\frac{1}{2}\int d^3x \left[-\frac{1}{2}(\d_\mu \Phi_0^a)(\d_\nu\Phi_0^b)g^{\mu\nu}Q_{ab}
  +\frac{1}{2}A_\mu^i M_{ij}^{\mu\rho}A_\rho^j+A_\mu^i (V_i^\mu -i W_i^\mu)\right]\cr
  &=&-\frac{1}{2}\int d^3x \left[(\d_\mu\Phi_0^a)(\d_\nu\Phi_0^b)g^{\mu\nu} Q_{ab}
  +(V-iW)_i^\mu {M^{-1}}_{\mu\rho}^{ij}(V-iW)^\rho_j\right].
\eea
We can couple to an external gauge field in the same way as before, by replacing derivatives with covariant derivatives either in $L_\mu^i$, or in the $\d_\mu\Phi$ itself. We can also add a vortex current to the action by replacing $A_{\mu}^i$ by $A_{\mu, {\rm smooth}}+A_{\mu,{\rm vortex}}$, with a corresponding equation of motion
\be
  A_{\mu, {\rm smooth}}+A_{\mu,{\rm vortex}}=-{M^{-1}}_{ij}^{\mu\rho}(V_j^\rho -iW_j^\rho)\;.
\ee
The new vortex current term in the action remains
\be
  \epsilon^{\mu\nu\rho}v_\mu^i (\d_\nu A^i_{\rho, {\rm vortex}}-\d_\rho A^i_{\nu,{\rm vortex}})
  \equiv v_\mu^i j^{\mu i}_{\rm vortex}.
\ee

\section{Some examples}
Having established, in some detail, the formal aspects of the construction, in this section we will illustrate explicitly 
how the nonabelian particle-vortex duality works in some important explicit cases.

\subsection{Semilocal (cosmic) strings}
The first case that we turn our attention to is that of the semilocal cosmic strings of 
\cite{Vachaspati:1991dz,Hindmarsh:1991jq,Gibbons:1992gt}. Cosmic strings are, of course, $(3+1)$-dimensional extensions of the usual $(2+1)$-dimensional abelian-Higgs vortices and have, over the years, played a not insignificant part in cosmological models of the early Universe and, in particular, the theory of structure formation \cite{Vilenkin:2000jqa}. {\it Semilocal} cosmic strings extend the usual $U(1)$ strings to solutions that exhibit an admixture of local and global symmetries that make them really the first examples of vortices with nonabelian structure. It also makes them a prime candidate to test our formulation of the duality.\\

\noindent
The model, as it was first formulated, has an abelian $U(1)$ gauge symmetry and a global $SU(2)$ symmetry. 
For ease of comparison with the general case, we will formulate it as a global $U(2)$ symmetry, 
a diagonal $U(1)$ subgroup of which is gauged. Besides the $U(1)$ gauge field $a_\mu$, the model contains a scalar
$\Phi=(\Phi^a)=\left(\Phi^1, \Phi^{2}\right)^{T}$, $a=1,2$, that transforms in the {\em fundamental} representation of the global flavor $SU(2)$. With a quartic Higgs-type potential for $\Phi$, the action takes the form
\be
  S = \int d^3x\left[-\frac{1}{2}|D_{\mu}\Phi|^{2} - \frac{\lambda}{4}\left(\Phi^{\dagger}\Phi - v^{2}\right)^{2} 
  - \frac{1}{4}f_{\mu\nu}f^{\mu\nu}\right].\label{cosmic}
\ee
Here the covariant derivative of $\Phi$ is the usual abelian one, $D_\mu \Phi=(\d_\mu -ie a_\mu)\Phi$ with a standard field strength $f_{\mu\nu}=\d_\mu a_\nu -\d_\mu a_\nu$. For comparison with our more general formulation above, note that this is nothing but the abelian case described in section 3.2 with the index $A$ being in a singlet, while 
the index $a$, is in the fundamental. In other words, $\Phi^a$ is coupled to an abelian external gauge field $a_\mu$. The ansatz in terms of the real field $\Phi_0^a$ and $L_\mu^i$ is therefore
\be
  \Phi^{a}=\Phi_0^a\, \exp\left(i \int dx^\mu L_\mu^i F_i\right), \;\;\; a=1,2;\;\; i=1,2,3,4.\label{phiphil}
\ee
The index $i$ is in the adjoint of the global $U(2)$, with $i=4$ corresponding to the identity, {\it i.e.} to the $U(1)$ that couples to the external gauge field, while $i=1,2,3$ correspond to the $SU(2)$ to be gauged for the duality procedure. As explained in our general discussion, in order to have the minimal coupling of the gauge field $a_\mu$ to $\Phi$ expressed by the replacement of $L^i_\mu$ with $\tilde L^i_\mu=L_\mu^i+a_\mu\,\delta_4^i$, we require $F_4=1$.\\

\noindent
The semi-local string solution of the theory is obtained from the axially symmetric $N$-vortex ansatz
\be
  a_\theta=\frac{v}{\sqrt{2}}\frac{n}{r}a(r);\;\;\; a_r=0;\;\;\;
  \Phi^a= v\, \varphi^a(r)\,e^{in\a_a}\;,
\ee
where $(r,\theta)$ are polar coordinates on the plane. The ansatz implies the condition that on the circle at infinity, $\a_2=\a_1+c$, with $c$ a constant. We can, without loss of generality, take $c=0$, in which case the phases of $\Phi^1,\Phi^2$ coincide and the  decomposition (\ref{phiphil}) required for the duality is satisfied on the $N$-vortex ansatz.\\

\noindent
The Hamiltonian of this system obeys a Bogomolnyi bound that saturates at the critical coupling 
$\beta\equiv \frac{2\lambda}{e^{2}} =1$, and where the second order equations of motion reduce to the first order 
BPS equations, 
\be
  \frac{d\varphi}{dr}=\frac{n}{r}(1-a)\varphi,\;\;\;\;
  \frac{da}{dr}=\frac{r}{n}(1-\varphi^2)\;,
\ee
with $\varphi(r)=\sqrt{(\varphi^1(r))^2+(\varphi^2(r))^2}$. These are the same as the BPS equations of the Nielsen-Olesen vortex, a fact that can be exploited to allow us to construct the semi-local vortex from the numerical solution for the former. On the decomposition ansatz (\ref{phiphil}), the action (\ref{cosmic}) becomes
\be
  S_{\rm original}=\int d^3x\left[-\frac{1}{2}(\d_\mu\Phi_0^a)^2-\frac{1}{2}(\Phi_0^a)^2g^{\mu\nu}\sum_{i,j=1}^4\tilde L_\mu^i  
  \tilde L_\nu^jE_{ij}
  -\frac{1}{4}f_{\mu\nu}^2- \frac{\lambda}{4}\left((\Phi_0^a)^2 - v^{2}\right)^{2}\right]\,,\label{originalcosmic}
\ee
where $E_{ij}=F_iF_j$ and $\Phi_{0}^{a} = v\varphi^{a}$. To dualize, we first gauge the model by replacing 
$\tilde L_\mu^i$ with 
\be
\tilde{\tilde{L}}_\mu^i=-i\Tr\left[T^ig^{-1}\left(\d_\mu-i\left(\sum_{\tilde i=1,2,3}A_\mu^{\tilde i}T_{\tilde i}+a_\mu^4 T_4\right)g\right)\right]\;,
\ee
then add a Lagrange multiplier term for $A_\mu^{\tilde i}$, obtaining the master action
\bea
  S_{\rm master}&=&\int d^3x \left[-\frac{1}{2}(\d_\mu\Phi_0^a)^2-\frac{1}{2}(\Phi_0^a)^2g^{\mu\nu}\sum_{i,j=1}  
  ^4\tilde{\tilde{L}}_\mu^i\tilde{\tilde{L}}_\nu^jE_{ij}
  -\frac{1}{4}f_{\mu\nu}^2\right.\cr
  &&\left.- \frac{\lambda}{4}\left((\Phi_0^a)^2 - v^{2}\right)^{2}+\epsilon^{\mu\nu\rho}\sum_{\tilde i=1,2,3}v_\mu^{\tilde i} F_{\nu  
  \rho}^{\tilde i}
  \right].
\eea
Eliminating the Lagrange multipliers $v^{\tilde i}$, we can put $A_\mu^i=0$ and return to the original action 
(\ref{originalcosmic}), which is equivalent to (\ref{cosmic}) on the decomposition ansatz. Gauge fixing 
$g=1$ for the $SU(2)$ part and eliminating the resulting $A_\mu^{\tilde i}$ instead, obtains the dual action
\bea
  S_{\rm dual}&=&\int d^3x\left[-\frac{1}{2}(\d_\mu \Phi_0^a)^2-\frac{1}{4}f_{\mu\nu}^2- \frac{\lambda}{4}\left((\Phi_0^a)^2   
  - v^{2}\right)^{2}\right.\cr
  &&\left.+A_\mu^{\tilde i}( V_{\tilde i}^\mu+M_{\tilde i 4}^{\mu\rho}a_\rho)+
  \frac{1}{2}a_\mu g^{\mu\rho}(\Phi_0^a)^2 a_\rho
  +\frac{1}{2}A_\mu^{\tilde i} M_{\tilde i\tilde j}^{\mu\rho}A_\rho ^{\tilde j}\right]\;,
\eea
where $A_\mu^{\tilde i}=-M_{\tilde i\tilde j}^{-1\mu\rho}(V_{\tilde j}^\rho+M_{\tilde i 4}^{\rho\sigma}a_\sigma)$, and $M$ and $V$ were defined in (\ref{mandv}). Finally then,
\bea
S_{\rm dual}&=&-\int d^3x\left[\frac{1}{2}(\d_\mu \Phi_0^a)^2+\frac{1}{4}f_{\mu\nu}^2+\frac{\lambda}{4}\left((\Phi_0^a)^2 - v^{2}\right)^{2}\right.\cr
&&\left.+\frac{1}{2}(V_{\tilde i}^\mu+M_{\tilde i 4}^{\mu\rho}a_\rho)M_{\tilde i\tilde j}^{-1\mu\nu}(V_{\tilde j}^\nu+M_{\tilde j 4}^{\nu\sigma}a_\sigma)
+\frac{1}{2}a_\mu g^{\mu\rho}(\Phi_0^a)^2 a_\rho\right].
\eea
Since the duality exchanges particles with vortices, the basic excitations of the dual field are the semilocal vortices. In this sense, the dual action encodes the full (perturbative) dynamics of semilocal strings. In principle then, phenomena like string condensation (as a seed for structure formation, for example) can be understood from this action.

\subsection{Theory with adjoint matter: modification of ${\cal N}=2$ $SU(2)$ SYM}
We now move on to a case with adjoint matter. We will start with a $(2+1)$-dimensional reduction of the familiar ${\cal N}=2$ $SU(2)$ supersymmetric Yang-Mills theory studied by Seiberg and Witten \cite{Seiberg:1994rs,Seiberg:1994aj} in $(3+1)$-dimensions.
The model has $SU(2)$ gauge fields $A_\mu=A_\mu^aT_a$, and adjoint scalars $\Phi=\Phi^aT_a$, together with their fermionic superpartners. Ignoring the fermions for this dicussion, the bosonic action in $(2+1)$ dimensions takes the form
\be
  S=-\int d^3x \Tr\left[\frac{1}{4}F_{\mu\nu}^2+|\tilde D_\mu\Phi|^2+\frac{g^2}{2}[\Phi^\dagger,\Phi]^2\right]\;,
\ee
where $\tilde D_\mu$ is an adjoint covariant derivative. The model also has a global $U(1)$ symmetry $\Phi\rightarrow e^{i\a}\Phi$. Dualizing along this direction, however, means that the generator left ungauged is abelian. Consequently, the resulting particle-vortex duality would be abelian too, with not much further to discuss. If, however, we turn off the adjoint gauge field, and instead gauge the $U(1)$ gauge field, the following action is obtained
\be
S=-\int d^3x \left[\frac{1}{4}F_{\mu\nu}F^{\mu\nu} +|D_\mu \Phi^a|^2+2g^2[\epsilon_{abc}\Phi^b\Phi^{\dagger c}]^2\right]\;,
\ee
where $D_\mu\Phi^{a}=(\d_\mu-ig A_\mu)\Phi^a$. Then, with the decomposition ansatz
\be
  \Phi^a=\Phi_0^a e^{i\int \sum_{i=1}^4L_\mu^i F_i^A\widetilde{T}_A}\;,
\ee
where $\Phi_0^a$ is a real field in the adjoint of $SU(2)$, $i=1,...,4$ with $i=4$ corresponding to the identity, 
and $\widetilde{T}_A$ is another adjoint representation of $SU(2)$, the action becomes
\bea
  S_{\rm original}&=&
  -\int d^3x\left[\frac{1}{4}F_{\mu\nu}F^{\mu\nu} +3((\d_\mu\Phi_0^a)^2+(\Phi_0^a)^2g^{\mu\nu}(L_\mu^i+A_\mu\delta_4^j)   
  (L_\nu^j+A_\mu \delta_4^j) 
  E_{ij})\right.\cr
  &&\left.+2g^2(\epsilon_{abc}\Phi_0^b\Phi_0^c)^2\right].
\eea
The master action for the duality with respect to the global $SU(2)$ is then 
\bea
S_{\rm master}&=&-\int d^3x\left[\frac{1}{4}F_{\mu\nu}F^{\mu\nu} +3((\d_\mu\Phi_0^a)^2+(\Phi_0^a)^2g^{\mu\nu}(\tilde{L}_\mu^i+A_\mu\delta_4^j) 
(\tilde L_\nu^j+A_\mu \delta_4^j) E_{ij})\right.\cr
&&\left.+2g^2(\epsilon_{abc}\Phi_0^b\Phi_0^c)^2+\epsilon^{\mu\nu\rho}\sum_{i=1}^3v_\mu^iF_{\nu\rho}^i\right].
\eea
The dual action is found to be
\bea
S_{\rm dual}&=&-\int d^3x\left[\frac{1}{4}F_{\mu\nu}F^{\mu\nu} +\frac{1}{2}(\d_\mu\tilde\Phi_0^a)^2+\frac{g^2}{18}(\epsilon_{abc}\tilde\Phi_0^b
\tilde\Phi_0^c)^2\right.\cr
&&\left.+\frac{1}{2}(V_{\tilde i}^\mu+M_{\tilde i 4}^{\mu\rho}A_\rho)M_{\tilde i\tilde j}^{-1\mu\nu}(V_{\tilde j}^\nu+M_{\tilde j 4}^{\nu\sigma}A_\sigma)
+\frac{1}{2}A_\mu g^{\mu\rho}(\tilde\Phi_0^a)^2 A_\rho\right]\;,
\eea
where $\tilde \Phi_0^a=\Phi_0^a/\sqrt{6}$ and the matrices $M$ and $V$ have the same definitions as before.

\subsection{Color-flavor locking and nonabelian vortices}
Finally, we consider the theory in which the original nonabelian vortex was constructed \cite{Auzzi:2003fs}. The model is one with matter in the fundamental of the gauge group with $N_f$ flavors, and whose vortices are nonabelian in the sense that they are not dynamically reducible to abelian solitons. The model considered in \cite{Auzzi:2003fs} is the low energy limit of an ${\cal N}=2$ supersymmetric $SU(3)$ gauge theory with $N_f$ flavors and equal quark masses. It has an $SU(2)\times U(1)$ gauge group, with field content consisting of the gauge fields $A_\mu^a$, $a=1,2,3$ and $A_\mu$; adjoint complex scalar superpartners, $a^a$ and $a$, of the gluons and their the corresponding fermions; bi-fundamental quark fields $q^{kA}$ and $\tilde q_{kA}$, in $SU(2)$ doublets (so that $k=1,2$) and in the fundamental of the $SU(N_f)$ flavor symmetry (so that $A=1,...,N_f$), as well as their corresponding fermions. The bosonic part of the low-energy action, considered in 2+1 dimensions, reads
\bea
  S&=&-\int d^3x \left[\frac{1}{4g_2^2}(F^a_{\mu\nu})^2+\frac{1}{4g_1^2}(F_{\mu\nu})^2+\frac{1}{g_2^2}|\nabla_\mu 
  a^a|^2\right.\cr
  &&\left.+\frac{1}{g_1^2}|\d_\mu a|^2
  +|D_\mu q^A|^2+|D_\mu \bar{\tilde q}^A|^2+V\left(a^a,a,q^A,\tilde q_A\right)\right].
\eea
The potential $V$ is, in general, some complicated admixture of F- and D-terms, but whose exact form is not relevant to our discussion. The covariant derivative $\nabla$ is in the adjoint of $SU(2)$, while  
\be
  D_\mu=\d_\mu-iA_\mu^a\frac{\tau^a}{2}-i\frac{1}{2\sqrt{3}}\,A_\mu.
\ee
The most relevant case for our discussion is $N_f=2$, so we will focus exclusively on this from now on. For this number of flavors, this model exhibits nonabelian vortex solutions, with an ansatz which manifests ``color-flavor locking". In other words, with the quarks $q^{kA},\tilde q_{kA}$ as matrices in the $2\times 2$ dimensional $(N_c,N_f)$ space, the solutions are diagonal. We will also take the adjoint scalars $a^a$ and $a$ to be trivial {\it i.e.}, taking their constant VEVs. As such, the vortex ansatz is
\bea
  q^{kA}&=&\begin{pmatrix} e^{in\theta}\phi_1(r) & 0\\ 0 & e^{ik\theta}\phi_2(r)\end{pmatrix},\cr
  A_i^3&=&-\eta \epsilon_{ij}\frac{x^j}{r^2}(n-k-f_3(r));\;\;\; A_i^1=A_i^2=0,\cr
  A_i&=&-\sqrt{3}\eta \epsilon_{ij}\frac{x^j}{r^2}(n+k-f(r))\;,
\eea
where $\eta=\pm $ is a sign and $\phi_1(r),\phi_2(r), f_3(r)$ and $f(r)$ are real functions that satisfy some BPS equations whose solutions will define the vortex profile.\\

\noindent
The full symmetry group of the theory is $SU(2)\times SU(2)\times U(1)$, out of which an $SU(2)\times U(1)$ is coupled to external gauge fields. In other words, we could well perform the duality on the global (flavor) $SU(2)$. In this case, we would 
need 7 $L_\mu^i$ for the 7 generators of the full symmetry group. However, if we were to gauge the whole $SU(2)$ using the prescription for a general representation (\ref{generalcov}), we would obtain a different 
covariant derivative than the one in used in the above model. The only possibility that allows us to have the nonabelian duality described in this article then, is to restrict to abelian covariant derivatives with respect to the overall $U(1)$ and the $U(1)$ defined by the $\tau_3\in SU(2)$. This means that, together with the $SU(2)$ to be gauged, we need 5 $L_\mu^i$'s in the decomposition ansatz. Since the ansatz for the vortex takes
\be
  q^{1A}=\begin{pmatrix} e^{in\theta}\\ 0\end{pmatrix};\;\;\;
  q^{2A}=\begin{pmatrix} 0\\e^{ik\theta} \end{pmatrix}\;,
\ee
we may as well write the quark fields as
\be
  q^{1A}=\Phi_0^1\, e^{i\int dx \sum_{i=1}^5 L_\mu^i F_i};\;\;\;
  q^{2A}=\Phi_0^2\, e^{i\int dx \sum_{i=1}^5 L_\mu^i F_i}\;,
\ee
with covariant derivatives
\be
  \tilde D_\mu q^{1A}=(\d_\mu-ia_\mu^{(1)})q^{1A};\;\;\;
  \tilde D_\mu q^{2A}=(\d_\mu-ia_\mu^{(2)})q^{2A}\;,
\ee
and where
\be
  a_\mu^{(1)}=A_\mu+A_\mu^3;\;\;\;
  a_\mu^{(2)}=A_\mu-A_\mu^3.
\ee
This is therefore consistent with both the vortex ansatz, and the general duality prescription described in this paper. 
The original action for the duality is then
\bea
  S_{\rm original}&=&-\int d^3x \left[\frac{1}{4g_2^2}(F^a_{\mu\nu})^2+\frac{1}{4g_1^2}(F_{\mu\nu})^2+
  \frac{1}{2}(\d_\mu\Phi_0^1)^2+\frac{1}{2}(\d_\mu\Phi_0^2)^2\right.\cr
  &&\left.+\frac{1}{2}(\Phi_0^1)^2\left(L_\mu^i-\delta_4^\mu a_\mu^{(1)}\right)
  \left(L_\nu^j-\delta_4^ja_\nu^{(1)}\right)g^{\mu\nu}
  E_{ij}\right.\cr
  &&\left.+\frac{1}{2}(\Phi_0^2)^2\left(L_\mu^i-\delta_5^\mu a_\mu^{(2)}\right)
  \left(L_\nu^j-\delta_5^ja_\nu^{(2)}\right)g^{\mu\nu}E_{ij}+V\right]\;,
\eea
and corresponds to the master action  
\bea
  S_{\rm master}&=&-\int d^3x \left[\frac{1}{4g_2^2}(F^a_{\mu\nu})^2+\frac{1}{4g_1^2}(F_{\mu\nu})^2+
  \frac{1}{2}(\d_\mu\Phi_0^1)^2+\frac{1}{2}(\d_\mu\Phi_0^2)^2\right.\cr
  &&\left.+\frac{1}{2}(\Phi_0^1)^2\left(\tilde L_\mu^i-\delta_4^\mu a_\mu^{(1)}\right)
  \left(\tilde L_\nu^j-\delta_4^ja_\nu^{(1)}\right)\,g^{\mu\nu}\,
  E_{ij}\right.\cr
  &&\left.+\frac{1}{2}(\Phi_0^2)^2\left(\tilde L_\mu^i-\delta_5^\mu a_\mu^{(2)}\right)
  \left(\tilde L_\nu^j-\delta_5^ja_\nu^{(2)}\right)\,g^{\mu\nu} \,
  E_{ij}+V
  +\epsilon^{\mu\nu\rho}\sum_{\tilde i=1,2,3}v^{\tilde i} F_{\mu\nu}^{\tilde i}\right].\cr
  &&
\eea
The dual action is found to be
\bea
  S_{\rm dual}&=&-\int d^3x \left[\frac{1}{4g_2^2}(F^a_{\mu\nu})^2+\frac{1}{4g_1^2}(F_{\mu\nu})^2+
  \frac{1}{2}(\d_\mu\Phi_0^1)^2+\frac{1}{2}(\d_\mu\Phi_0^2)^2\right.\cr
  &&\left.+A_\mu^{\tilde i}( V_{\tilde i}^\mu+M_{\tilde i 4}^{\mu\rho}a_\rho^{(1)}+M_{\tilde i 5}^{\mu\rho}a_\rho^{(2)})+
  \sum_{k=1,2}\frac{1}{2}a_\mu^{(k)} g^{\mu\rho}(\Phi_0^k)^2 a_\rho^{(k)}
  +\frac{1}{2}A_\mu^{\tilde i} M_{\tilde i\tilde j}^{\mu\rho}A_\rho ^{\tilde j}\right]\cr
  &=&-\int d^3x \left[\frac{1}{4g_2^2}(F^a_{\mu\nu})^2+\frac{1}{4g_1^2}(F_{\mu\nu})^2+
  \frac{1}{2}(\d_\mu\Phi_0^1)^2+\frac{1}{2}(\d_\mu\Phi_0^2)^2\right.\cr
  &&\left.+\frac{1}{2}(V_{\tilde i}^\mu+M_{\tilde i 4}^{\mu\rho}a_\rho^{1)}+M_{\tilde i 5}^{\mu\rho} A_\rho^{(2)})
  M_{\tilde i\tilde j}^{-1\mu\nu}(V_{\tilde j}^\nu+M_{\tilde j 4}^{\nu\sigma}a_\sigma^{(1)}+M_{\tilde j 5}^{\nu\sigma} a_\sigma^{(2)})\right.\cr
&&\left.+\frac{1}{2}\sum_{k=1,2}a_\mu^{(k)} g^{\mu\rho}(\Phi_0^k)^2 a_\rho^{(k)}\right]\;,
\eea
with the same definition for $M$ and $V$, but in the definition of the matrix $M$ we now replace $(\Phi_0^a)^2$ with $(\Phi_0^1)^2+(\Phi_0^2)^2$.

\section{Conclusions}
This article is an elaboration on a previous paper in which we first defined a new nonabelian particle-vortex duality \cite{Murugan:2015boa}. The low-
dimensional duality that we describe in detail here not only extends the, by now, well-known $U(1)$ particle-vortex duality to larger gauge groups, but it also 
generalizes to 2+1 dimensions, nonabelian T-duality on the string
worldsheet. It acts on a theory with nonabelian global symmetry, $G$, as well as an extra local symmetry, 
and with matter in an arbitrary representation of the gauge group. As concrete illustrations of how the duality works in practice, we have studied three different 
theories that all exhibit nonabelian vortices of some form: semilocal cosmic strings, a modification of  ${\cal N}=2$ SYM gauge theory (Seiberg-Witten model) 
reduced to 2+1 dimensions, and the original color-flavor locked nonabelian vortices. In each of these cases, we showed how our vortex ansatz could be used 
to be used to implement the duality $S_{\mathrm{original}}\to S_{\mathrm{master}} \to S_{\mathrm{dual}}$ at the level of the on-shell action.\\

\noindent
There remain a number of important points that we have not addressed yet. Foremost among these is the issue of topology. It is well known that topological 
issues make implementing nonabelian T-duality at the level of the full path integral notoriously difficult \cite{Giveon:1993ai}. Given that our construction 
hinges on a lifting of nonabelian T-duality from 2 to 3 dimensions, and that 3-dimensional spacetime is replete with topological nuances (winding numbers, 
Chern-Simons theory, etc.), we expect that such issues, while beyond the scope of this paper, will no doubt play a role in a full understanding of the duality. 
Another, perhaps related, point that deserves further clarification is that, unlike full nonabelian T-duality which acts on the full theory, particle-vortex duality 
acts only on-shell and maps only a small sector of the one theory into the other. We anticipate that, clarifying 
the domain and range of the duality map will also shed some much needed light on the moduli space of vortex zero-modes. Finally, nonabelian vortices of the 
kind studied in this paper are closely related to the nonabelian monopoles (of the higher dimensional Seiberg-Witten like theory) on which they end. In this 
sense, the particle-vortex duality we study here should be intimately related to electric-magnetic duality in the high dimensional theory. How exactly, remains 
to be understood.\\

\noindent
That particle-vortex duality is an extremely powerful tool with many potential applications is undisputed. In fact, the abelian version of the duality has already 
proven very useful in the analysis of the quantum Hall effect \cite{Burgess:2000kj}. Much of this utility comes from the structure of the modular group 
associated to the duality transformation, itself inherited from the symmetry group of the worldsheet T-duality group. Worldsheet nonabelian T-duality 
transformations no longer form a group, so certainly, we anticipate that at least some of this utility is lost. Nevertheless, it would be very interesting to see how 
much of this structure is still preserved.\\

\noindent
To close, one further interesting, and quite topical, direction for development is the following. Theories with Chern-Simons gauge fields, including so-called 
Chern-Simons-matter theories, like the ABJM model, are known to support {\it anyons}, quasiparticles that obey neither bosonic nor fermionic statistics 
\cite{Kawamoto:2009sn}. In fact, a nonrelativistic limit of an abelian reduction of the ABJM model \cite{Lopez-Arcos:2015cqa} leads to a theory that can be 
used to describe anyons in a harmonic trap \cite{Doroud:2015fsz}. It is intriguing to speculate that the non-Abelian particle-vortex duality studied here, 
embedded into a non-Abelian Chern-Simons gauge theory like the ABJM model \cite{Murugan:2014sfa}, might be relevant for a holographic description of the 
nonabelions of \cite{Moore:1991ks}. These are quasiparticle excitations of fractional quantum Hall systems that obey nonabelian braiding statistics and play a 
central role in the study of topological quantum matter. But this is another story for another year.\\

\noindent
\section{Acknowledgements}
We thank Thiago Ara\'{u}jo, Leo Pando Zayas, Fernando Quevedo and Jonathan Shock for discussions.
The work of HN is supported in part by CNPq grant 301709/2013-0 and FAPESP grants 2013/14152-7 and 2014/18634-9.
JM acknowledges support from  the National Research Foundation (NRF) of South Africa under it CPRR program under grant number 87677.

\appendix

\section{General representation}
In this appendix we present some comments on the most general case, with the scalar $\Phi$ in an arbitrary representation. 
As before, the representation must split into a product of representations labelled by $a$ and $A$. Now however, to preserve generality, we cannot use generators (denoted earlier by $T_a$ and $\tilde T_A$), which have indices in the adjoint representation. Instead, we will make an analogous ansatz, without these matrices, 
\be
  \Phi^{a, A}=\Phi_0^a e^{i \int dx^\mu L_\mu^i F_i^A},
\ee
and note that there should be no sum over $A$ when expanding the exponential into products. At first sight, this might seem strange, since one would think that a product of fields in some representation, should be a matrix product with a corresponding sum over indices. Formally, though, one can simply say that an object $f^A$ defined, such that
$\Phi^{a, A}=\Phi_0^a f^A$, is written (and treated) as an exponential, say $(e^{i\int dx^\mu L_\mu^i F_i})^A$.\\ 

\noindent
To write an appropriate kinetic term for $\Phi$, we compute,
\be
  \d_\mu \Phi_0^a=\left[(\d_\mu\Phi_0^a)+i\Phi_0^a\, L_\mu^i\, F_i^A\right]\, e^{i\int dx^\mu L_\mu^i F_i^A}\;,
\ee
leading to the kinetic term for the ungauged case
\be
\sum_{a,A}|\d_\mu \Phi^{aA}|^2=N(\d_\mu\Phi_0^a)^2+(\Phi_0^a)^2 L_\mu^i L_\nu^jg^{\mu\nu}F_i^A F_j^A\equiv N[(\d_\mu\Phi_0^a)^2+(\Phi_0^a)^2L_\mu^i L_\nu^j
g^{\mu\nu}E_{ij}]\;,
\ee
i.e., the same form in terms of components as in the case of the adjoint representation. In order to gauge the theory, we need a covariant derivative for the gauge field. We can make the derivative in $L_\mu^i$ covariant, in the same way for the 
duality gauging with $A_\mu^i$, and for the gauging with the external field $a_\mu$, thus replacing $L_\mu^i$ with $\tilde{\tilde{L}}_\mu^i$. 

Or, since we would like to have a covariant derivative on the whole $\Phi^{aA}$ field in order to define finite energy vortices, we could define
\be
(D_\mu\Phi)^{aA}=(\d_\mu \Phi_0^a\delta_C^A+\Phi_0^a i L_\mu ^i F_i^C\delta^A_C +\Phi_0^a i A_\mu^M{(T_M)^A}_C)e^{i\int dx^\mu L_\mu^j F_j^C}.
\ee 
This assumes that the index $A$ on $F_i^A$ is gauged, so a certain identification of $i$ and $A$ indices, at least for the part that is gauged. 
Otherwise, the same replacement of $L_\mu^i$ with $\tilde{\tilde{L}}_\mu^i=L_\mu^i+a_\mu^i$ can be done in the exponent, 
\be
(D_\mu\Phi)^{aA}=(\d_\mu \Phi_0^a\delta_C^A+\Phi_0^a i L_\mu ^i F_i^C\delta^A_C)e^{i\int dx^\mu (L_\mu^j+a_\mu^j) F_j^C}.\label{generalcov}
\ee 
Note that $\Phi_0^a$ is real, so the gauging doesn't affect its index, $a$.
Besides the above possibilities, we have the cases of the adjoint and the $U(1)$ gaugings, already discussed. 
Which of the gaugings is chosen depends on the model.

\bibliography{PartVort}
\bibliographystyle{utphys}

\end{document}